\documentclass[fleqn,usenatbib]{mnras}
\usepackage{newtxtext,newtxmath}
\usepackage[T1]{fontenc}
\usepackage{ae,aecompl,graphicx,amsmath,comment,epsfig,psfrag}

\def\E33{\overline{E}_{33}}
\title[Constraining FRB properties]{Constraining the Fast Radio Burst (FRB) properties using the joint distributions of dispersion measure and fluence of the events detected at Parkes, ASKAP, CHIME and UTMOST}

\author[Bhattacharyya et al.] {Siddhartha Bhattacharyya$^1$\thanks{siddhartha@phy.iitkgp.ac.in} 
and Somnath Bharadwaj$^{1,2}$ \\ 
$^1$Department  of Physics, Indian Institute of Technology, Kharagpur, India \\ 
$^2$Centre for Theoretical Studies, Indian Institute of Technology, Kharagpur, India \\}

\begin{document}
\label{firstpage}
\pagerange{\pageref{firstpage}--\pageref{lastpage}}
\maketitle

\begin{abstract}
The Parkes, ASKAP, CHIME and UTMOST telescopes, which have all detected FRBs, each works at a different frequency and has a different detection criteria.  Using simulations, we have combined the constraints from all four telescopes to identify an allowed range of model parameters $(\alpha, \E33)$ for the FRB source population. Here  $\alpha$ is   the  spectral index and $\E33$  is the  mean FRB energy  in units of $10^{33} \, {\rm J}$ across   a $2128 - 2848\; {\rm MHz}$ band in the FRB rest frame.  We have considered several different FRB energy distributions, and also different scenarios for the scattering pulse broadening, the event rate density variation with $z$   and the host dispersion measure. We find that in all cases,  the common allowed region includes the range  $-3.9\leq\alpha\leq-1.3$ and $0.42\leq\E33\leq1$.  In all case, large values $\alpha > 4$ and $\E33 >60$ are ruled out. 
Considering the allowed  $(\alpha, \E33)$ parameter range, we predict that CHIME is unlikely to detect an FRB with  extra-galactic dispersion measure $(DM_{Ex})$ exceeding $3700\,{\rm pc\,cm}^{-3}$.  A substantially larger $DM_{Ex}$ in the large FRB sample anticipated  from CHIME would falsify the assumptions of the present analysis. Our analysis is expected to yield tighter parameter constraints with the advent of more FRB data.
\end{abstract}

\begin{keywords}
transients: fast radio bursts, scattering.
\end{keywords}

\section{Introduction}\label{sec:1}
Fast radio bursts (FRBs) are short duration ($\sim{\rm ms}$),highly energetic ($\sim 10^{31}-10^{34}\,{\rm J}$), dispersed radio pulses first detected at Parkes \citep{lorimer07}.  These have been detected at Parkes, ASKAP, CHIME, UTMOST, Arecibo, GBT, DSA-10 and Pushchino, of these several FRBs are found to repeat. The details of FRB observations are summarized in Table~\ref{tab:summary}. Several studies \citep{Palaniswamy18,caleb18,lu19} indicate  that the FRBs which are not found to repeat (non-repeating FRBs) form a different FRB  population from  the repeating FRBs. We have only considered the non-repeating FRBs for the entire analysis presented here. 
The observed dispersion measures (DMs), which are $\sim5$ to $20$ times in excess of the DMs of the Milky Way, strongly suggests that the FRBs are extragalactic in origin.  It has been possible to localise only a few FRB events \citep{chatterjee17,bannister19,ravi19,fedorova19a,prochaska19,marcote20,macquart20},  and direct redshift estimates are not available for most of the  FRB events detected till date. For most detected FRBs we only have redshifts inferred from the observed dispersion measures, these are rather uncertain because the dispersion measure contributed  by the host galaxy  is largely unknown for each event.

Several models have been proposed to explain the FRB source emission mechanism,  but none of them gives us a clear picture at present. For example, it has been proposed that FRBs may originate from the collapse of compact objects like neutron stars or white drawfs.
A recent detection of a double peaked radio burst \citep{chime20} from  the Galactic magnetar SGR 1953+2154 indicates that such sources may be the origin of at least  some  of the FRBs. Several authors 
\citep{ridnaia20,li20,tavani20} have reported detecting  X-ray bursts  corresponding   to the same event.
The reader is refereed to \citet{platts18} for a summary of the different FRB models prescribed to date. Several attempts have been reported to constrain the spectral index $\alpha$ of FRBs without giving a complete picture of the emission mechanism. Considering the $23$ FRBs detected at ASKAP, \citet{macquart19} have  determined the mean spectral index $\alpha=-1.6^{+0.2}_{-0.3}$ from the estimated values of $\alpha$ ($F_{\nu}\propto\nu^{\alpha}$) which are found to range from $-7.5$ to $+7$ across the sample. However, \citet{farah19} have showed that the detection of five new FRBs at UTMOST do not agree with the steep spectral index estimated by \citet{macquart19} and the FRBs spectra may turnover at $1\,{\rm GHz}$. In a few cases the non-detection of FRBs also give limits on the spectral index. \citet{houben19} have proposed the lower limit $\alpha>-1.2\pm-0.4$ for the spectral index of FRBs based on the fact that the FRB $121102$ was detected at $1.4\,{\rm GHz}$ and not simultaneously detected at $150\,{\rm MHz}$. 
Considering the non-detection of FRBs towards the Virgo cluster, \citet{agarwal19} have proposed the upper limit $\alpha<1.52$ for the spectral index of FRBs with $68\%$ confidence.  \citet{hardy17} have  searched for the  optical counterpart of FRB $121102$  jointly using the  high-speed optical camera ULTRASPEC and the Effelsberg radio  telescope, however  they do  not find any event, which places an upper limit $\alpha\leq-0.2$ for the spectral index of this FRB. A few FRB simulations and theoretical modelling have also been reported to constrain the spectral property of FRBs. 
\citet{ravi18} have modelled the FRB spectrum using a broken power law to explain the dearth of FRB detection at low frequencies, i.e. below $200\,{\rm MHz}$ \citep{rowlinson16,sokolowski18}. The reader is referred to \citet{james19} for further discussion on the spectral index of FRBs.

We present here a methodology which can be used to constrain the properties of the FRB population using the FRBs observed at the different telescopes. 
The total observed dispersion measure ($DM$) of any detected FRB is a sum of  three distinct contributions  
\begin{equation}
    DM=DM_{\rm MW}+DM_{\rm IGM}+DM_{\rm Host}/(1+z)
    \label{eq:DM}
\end{equation}
with $DM_{\rm MW}$, $DM_{\rm IGM}$ and  $DM_{\rm Host}$ arising from the Milky Way, the inter-galactic medium  and the FRB host galaxy  respectively. We use the 
NE$2001$ model \citep{cordes02} to estimate $DM_{\rm MW}$ which depends on the observed galactic latitude and longitude of the FRB event. 
 We use
\begin{equation}
    DM_{\rm Ex}=DM-DM_{\rm MW} \,.
    \label{eq:DMex}
\end{equation}
 to estimate $DM_{\rm Ex}$ which quantifies the  extragalactic (EX) contribution to the observed  $DM$. 
 Considering $DM_{\rm Ex}$, both   $DM_{\rm IGM}$ and $DM_{\rm Host}$ are unknown. It is possible to calculate   $DM_{\rm IGM}$  \citep{ioka03} assuming a homogeneous and completely ionized IGM,  if the FRB redshift $z$ is known. Typically it is not possible to directly measure the redshifts of the FRB events. Further, there are no direct estimates of $DM_{\rm Host}$ which is the $DM$ within the  
host galaxy. The value of $DM_{\rm Host}$ is likely to vary from galaxy to galaxy, and this will also depend on the location of  the FRB within the host galaxy. Quite often one assumes a plausible value for $DM_{\rm Host}$ and uses the resulting   $DM_{\rm IGM}$ to infer the FRB redshifts.  For our analysis we consider the extragalactic dispersion measure $DM_{\rm Ex}$, the fluence $F$ and the pulse width $w$ which can all be directly determined from FRB observations. For a given telescope, we quantify the observed FRB distribution using the joint cumulative distribution $C_{O}(DM_{\rm Ex},F)$ which is as the fraction of FRBs having extragalactic dispersion measure less than $DM_{\rm Ex}$ and fluence less than $F$. 

In this paper we present a methodology  to use the observed joint cumulative distribution $C_{O}(DM_{\rm Ex},F)$ to constrain the intrinsic properties of the FRBs as quantified by the the spectral index $\alpha$ and energy $E$ (defined later) of the FRB population. To keep the analysis simple, we have assumed  that the entire FRB population can be characterized by a single value of $\alpha$ which we constrain using the FRB observations. There is a considerable evidence that the non-repeating FRBs cannot be characterized by a single value of energy and it is necessary to consider a spread in the energy value. In this work for the energy $E$, we consider a Schechter luminosity function whose parameters we constrain using FRB observations. The methodology is presented in Section~\ref{sec:3} of this paper. The methodology presented here is quite general, and it can be extended to consider other parameterizations of the FRB population. Considering any particular model and parameter values for the FRB population, we calculate the predicted joint cumulative distribution  $C_{M}(DM_{\rm Ex},F)$ and use the Kolmogorov-Smirnov (KS) test to  determine whether our model predictions are consistent with the observed joint cumulative distribution $C_{O}(DM_{\rm Ex},F)$. 
 
The two dimensional  KS test is not as straight forward as the one dimensional KS test, and  following \citet{peacock83} we have used simulations to estimate the model prediction  $C_{M}(DM_{\rm Ex},F)$ and also to quantify the statistical significance of the deviations between $C_{M}(DM_{\rm Ex},F)$  and $C_{O}(DM_{\rm Ex},F)$. This allows us to rule out regions of the FRB parameter space, and thereby constrain the  intrinsic properties of the FRB population. 

We have considered the FRBs detected at four different telescopes namely Parkes, ASKAP, CHIME and UTMOST. The FRB distribution from each telescope was analyzed  separately to constrain the FRB parameters. 
Now we discuss the FRB observations at the four telescopes mentioned above along with other FRB detections. A total of $106$ FRBs had been reported around the beginning of 2020, these  are available in the online platform\footnote{http://frbcat.org/}. Of these $106$ FRBs, $95$ FRBs are non-repeating and $11$ FRBs are repeating. These $106$ FRB events are summarized in Table~\ref{tab:summary}. In this analysis we have considered $82$ non-repeating FRBs detected at Parkes, ASKAP, UTMOST and CHIME, and they are shown in Table~\ref{tab:1}. This is $77\%$ of the total $106$ FRBs summarized in Table~\ref{tab:summary}.
\begin{table}
    \caption{The number of non-repeating and repeating FRBs detected at ASKAP, Parkes, CHIME, UTMOST, Pushchino, Arecibo, GBT and DSA-10 with references as follows. 1. \citet{bannister17}, 2. \citet{shannon18}, 3. \citet{macquart18}, 4. \citet{bhandari19}, 5. \citet{qiu19}, 6. \citet{bannister19}, 7. \citet{prochaska19}, 8. \citet{agarwal19}, 9. \citet{petroff16}, 10. \citet{petroff17}, 11. \citet{keane16}, 12. \citet{ravi16}, 13. \citet{bhandari17}, 14. \citet{shannon17}, 15. \citet{price18}, 16. \citet{osolowski18a}, 17. \citet{osolowski18b}, 18. \citet{osolowski18c}, 19. \citet{bhandari18}, 20. \citet{boyle18}, 21. \citet{chime19a}, 22. \citet{chime19b}, 23. \citet{chime19c}, 24. \citet{caleb17}, 25. \citet{farah18a}, 26. \citet{farah18b}, 27. \citet{farah18c}, 28. \citet{farah19}, 29. \citet{fedorova19a}, 30. \citet{spitler14}, 31. \citet{patel18}, 32. \citet{masui15}, 33. \citet{ravi19}} 
    \centering
    \begin{tabular}{cccc}
    \hline
    Telescope & Number of & Number of & Reference \\  
     Name & Non-Repeating FRBs & Repeating FRBs & \\ 
    \hline
    ASKAP & $31$ & $-$ & $1-8$ \\
    Parkes & $29$ & $-$ & $9-19$\\
    CHIME & $11$ & $9$ & $20-23$ \\
    UTMOST & $11$ & $-$ & $24-28$ \\
    Pushchino & $11$ & $-$ & $29$ \\
    Arecibo & $-$ & $2$ & $30,31$ \\
    GBT & $1$ & $-$ & $32$ \\
    DSA$-10$ & $1$ & $-$ & $33$ \\
    \hline
    \end{tabular}
    \label{tab:summary}
\end{table}

Parkes and ASKAP both have parabolic reflectors of diameter $64\,{\rm m}$ and $12\,{\rm m}$ respectively, whereas CHIME and UTMOST both have cylindrical reflectors of dimension $20{\rm m}\times 100{\rm m}$ and $11.6{\rm m}\times 778{\rm m}$ respectively. Parkes and ASKAP both are operating at $L$ band, whereas CHIME and UTMOST both are operating in UHF band. The frequency range of these four telescopes are tabulated in Table~\ref{tab:1}. 
The telescope can only detect an FRB if its signal to noise ratio $(S/N)$  exceeds  the threshold $(S/N)_{\rm th}$ value required for a detection \citep{petroff16}. The quantity $(S/N)_{\rm th}$ translates to a limiting fluence $F_l$ which is given by
\begin{equation}
    F_l=[\Delta S]_{\rm 1\,ms}\times(S/N)_{\rm th}\times 1\,{\rm ms}
    \label{eq:Fl}
\end{equation}
where $[\Delta S]_{\rm 1\,ms}$ is the rms noise of the telescope for $1\,{\rm ms}$ of observation. 
An FRB is detected only if $F\times \sqrt{1{\rm ms}/w}\ge F_l$ \citep{bera16}. Considering $(S/N)_{\rm th}=10$, the values of $F_l$ for Parkes, ASKAP, CHIME and UTMOST are tabulated in the table \ref{tab:1}.  
\begin{table*}
    \caption{Considering the four telescopes Parkes, ASKAP, CHIME and UTMOST, this shows the respective frequency range, limiting fluence $F_l$, extragalactic dispersion measure $DM_{\rm Ex}$ range,  fluence $F$ range and observed pulse width $w$ range. The redshifts $z$ are inferred  assuming  $DM_{\rm host}=0$,  and the inferred redshifts would be smaller if we were to consider a non-zero $DM_{\rm host}$ contribution.}
    \centering
    \begin{tabular}{ccccccc}
    \hline
    Telescope & Frequency range & $F_l$ & $DM_{\rm Ex}$ range & $z$ & $F$ range & $w$  range\\
    Name & $({\rm MHz})$ & $({\rm Jy\,ms})$ & $({\rm pc\,cm^{-3}})$ & range & $({\rm Jy\,ms})$ & $({\rm ms})$ \\
    \hline
    Parkes & $1157-1546$ & $0.5$ & $137.8-2583.1$ & $0.14-2.63$ & $0.56-51.3$ & $0.4-24.3$ \\
    ASKAP & $1129-1465$ & $4.1$ & $76.1-951.7$ & $0.08-0.92$ & $8-420$ & $1.2-9$ \\
    CHIME & $400-800$ & $0.64$ & $78.68-978.8$ & $0.09-0.94$ & $9-50$ & $0.1-1.3$ \\
    UTMOST & $827-859$ & $3.25$ & $139.8-1894$ & $0.15-1.86$ & $6.7-95$ & $0.4-26$ \\
    \hline
    \end{tabular}
    \label{tab:1}
\end{table*}

\begin{figure*}
\centering
\includegraphics[scale=0.7]{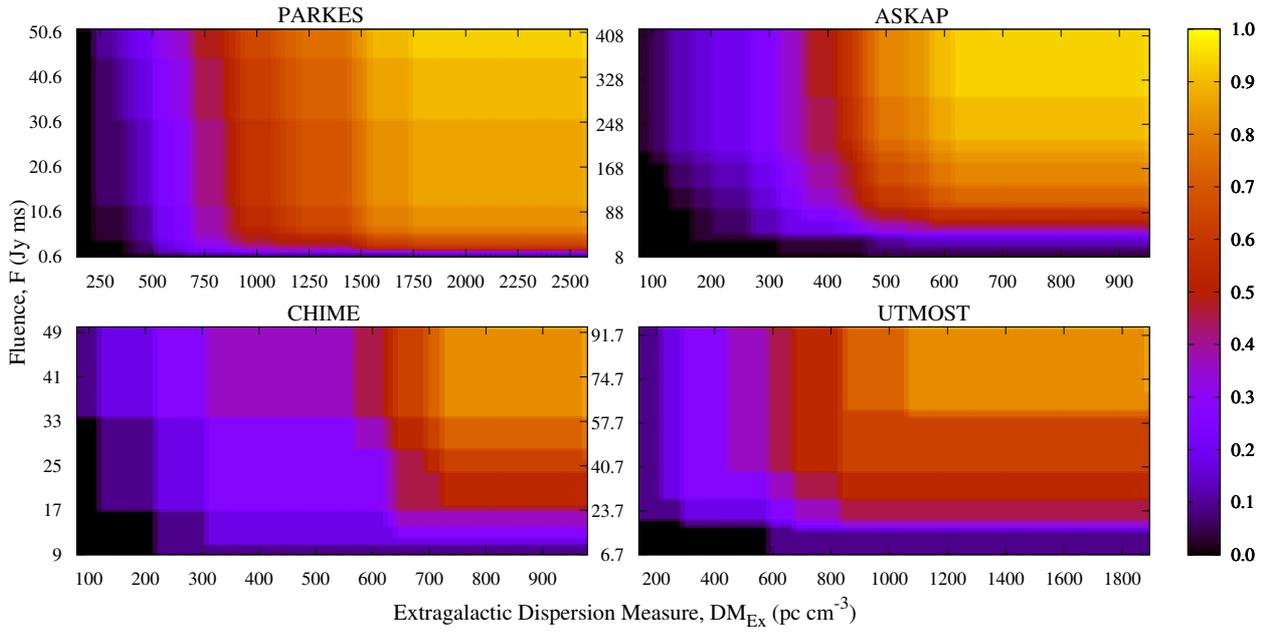}
\caption{The observed joint cumulative distribution $C_{O}(DM_{\rm Ex},F)$ of FRBs detected at Parkes, ASKAP, CHIME and UTMOST.}
\label{fig:2.1}
\end{figure*}

Each observed FRB is quantified by its  extragalactic dispersion measure $DM_{\rm Ex}$, fluence $F$ and pulse width $w$. For each telescope we have quantified the observed FRB distribution using the  joint cumulative distribution $C_{O}(DM_{\rm Ex},F)$  which gives the fraction of FRBs having extragalactic dispersion measure less than $DM_{\rm Ex}$ and fluence less than $F$.  
The different panels of Figure \ref{fig:2.1} shows $C_{O}(DM_{\rm Ex},F)$ for the FRBs detected at Parkes, ASKAP, CHIME and UTMOST respectively. Note that the observed $DM_{\rm Ex}$ and  $F$ range differs from telescope to telescope and these are tabulated in  
Table~\ref{tab:1} which also shows the $w$ range. Given the variation between the different telescopes, it is not very meaningful to directly compare the different  $C_{O}(DM_{\rm Ex},F)$  shown here. Later in this paper we  compare these with model predictions for the individual telescopes, and use this to constrain the model parameters. 

\begin{figure}
    \centering
    \includegraphics[width=\columnwidth]{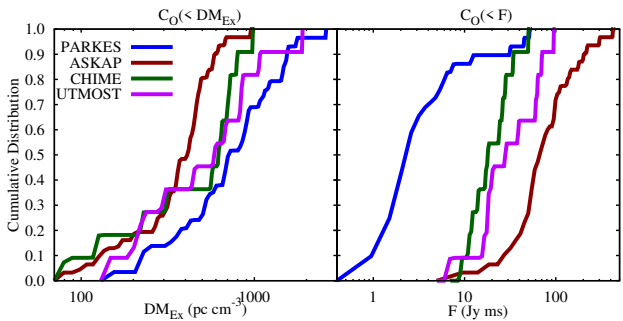}
    \caption{The observed cumulative distribution of extragalactic dispersion measure $C_O(DM_{\rm Ex})$ (left) and fluence $C_O(F)$ (right) of FRBs detected at Parkes (blue), ASKAP (red), CHIME (green) and UTMOST (magenta).}
    \label{fig:2.2}
\end{figure}

It is relatively easier to visualise one dimensional (1D) cumulative distributions  $C_O(<x)$ separately for $DM_{\rm Ex}$ and $F$ as shown in the left and right panels of Figure~\ref{fig:2.2} respectively. Each of these is obtained by collapsing $C_{O}(DM_{\rm Ex},F)$ along one of the directions. We notice that the 1D cumulative distributions for $DM_{\rm Ex}$ ($C_O(<DM_{\rm Ex})$)  are  somewhat similar for the different telescopes whereas these are considerably different for  $F$ ($C_O(<F)$).  The FBs detected at Parkes extend to the lowest fluence and highest $DM_{\rm Ex}$ which may be interpreted as the highest redshift.

A brief outline of the paper as follows. section~\ref{sec:3} presents the simulations and statistical analysis used in this work. The resulting constraints on the FRB parameters are presented in Section~\ref{sec:4}, and Section~\ref{sec:5} presents the summary and conclusions.

\section{Methodology}\label{sec:3}
We define the energy $E$ and spectral index $\alpha$ of an FRB through the specific energy density $E_{\nu}=E\,\phi(\nu)$. Here  $\phi(\nu)$ the emission frequency  profile  is assumed to be a power law $\phi(\nu)=K\nu^{\alpha}$ across the entire $\sim 1 \, {\rm GHz}$ spectral range of our interest.
The normalisation constant $K$ is determined  by $\int_{\nu_a}^{\nu_b}\phi(\nu)d\nu=1$. 
Following \citet{bera16}, the frequency profile $\phi(\nu)$ is normalised with respect to the reference event FRB $110220$ \citep{thornton13}, and   $\nu_a=2128\;{\rm MHz}$ and $\nu_b=2848\;{\rm MHz}$ correspond to the limits of the Parkes observing band redshifted to the estimated rest frame $(z=0.81)$ of FRB $110220$.  We note that the redshift of this  FRB has been estimated assuming  a null contribution from the host galaxy to the dispersion measure. Further,  the  above mentioned frequency  limits are obtained by considering the Parkes observational band  at the time when FRB $110220$ was detected.

FRBs observations at different redshifts and   different telescopes each probe  the energy emitted in a different range of frequencies. However, in order to compare the intrinsic properties of different FRB events it is desirable to consider the energy emitted in a fixed  frequency range.  In our analysis  $E$ refers to  the total energy emitted in the frequency interval $\nu_a=2128\,{\rm MHz}$ to $\nu_b=2848\,{\rm MHz}$ in the rest frame of the FRB . 
Considering an FRB detected  at redshift $z$ with a given telescope, we can use \citep{bera16} the energy $E$ to predict  the observed  fluence $F$ associated with each event through
\begin{equation}
F = \frac{\overline{\phi}(z)\,B(\vec{\theta})}{4\pi r^2}\, E
\label{eq:fluence} 
\end{equation}
where $B(\vec{\theta})$ is the telescope's beam pattern evaluated at the FRB's angular position $\vec{\theta}$, and  $\overline{\phi}(z)$ is the frequency profile  $\phi(\nu)$ averaged over the telescope's observing bandwidth redshifted to the FRB's rest frame. As mentioned earlier, we have assumed $\phi(\nu)$ to be a power law across the entire $\sim 1 \, {\rm GHz}$ spectral range of our interest. The function $\overline{\phi}(z)$ above accounts for the fact that for each FRB we actually measure the energy emitted in a different frequency range, and   
the relation between $F$ and $E$  depends on $\alpha$ through  $\overline{\phi}(z)$.

In addition to the energy $E$ and spectral index $\alpha$, we also associate an intrinsic pulse width $w_i$ with every FRB event. This is the emitted pulse width in the rest frame of the FRB. 
The value of  $w_i$ depends on the source dimension, and it may vary from event to event. We do not have a definite  picture about $w_i$ at present, and for our analysis we assume that all the FRBs have the same value $w_i=1\,{\rm ms}$. As discussed later, we have also considered $w_i=0.5\,{\rm ms}$ and  $2\,{\rm ms}$, and we find that this variation of $w_i$ does not significantly affect our results.
The entire formalism used here  closely  follows \citet{bera16} to which the reader is referred for further details.

We have quantified  the FRB population across different redshifts by using the function $n(E,z)$ which is defined through  $dN=n(E,z)\,dE$, where $dN$ is the FRB event rate per unit comoving volume  with energy in the range $E$ to $E+dE$.
The energy distribution of FRBs is presently  unknown. While  there are indications that the FRBs have a spread in energies, it is reasonable to assume that the FRBs have a characteristic energy scale and the distribution  falls off rapidly at energies larger than the characteristic energy. This motivates us to adopt the  Schechter luminosity function \citep{schechter76} as a possible energy distribution model for the FRBs. We have modelled  the FRB energy distribution function using  
\begin{equation}
    n(E,z)=%
    \begin{cases}
    \frac{n_0(z)}{\overline{E}}\;\exp\left(-\frac{E}{\overline{E}}\right) & {\rm for}\;\gamma=0 \\
    \frac{n_0(z)}{\overline{E}}\left(\frac{\gamma}{\Gamma(1+\gamma)}\right)\left(\frac{\gamma E}{\overline{E}}\right)^{\gamma} \exp\left(-\frac{\gamma E}{\overline{E}}\right) & {\rm for}\;\gamma>0
    \end{cases}
    \label{eq:sch}
\end{equation}
where $\overline{E}$ is the mean energy of the population, $\gamma$ is the exponent and $\Gamma(n)$ is the Gamma function. The FRB distribution is normalised in such a way that $n_0(z)$ gives  the FRB event rate per unit comoving volume. For $\gamma<0$ the FRB distribution increases with decreasing $E$ and one has to introduce  a lower energy cut-off to make the distribution normalizable. Here we have restricted $\gamma\geq0$ to avoid this.  
For $\gamma=0$ the FRB distribution is constant up to $E\leq\overline{E}$ beyond which it decreases with increasing $E$. For $\gamma>0$ the FRB distribution increases with $E$ for $E<\overline{E}$, it peaks at $E=\overline{E}$ and then decreases with $E$. The width of the FRB population decreases with increasing $\gamma$, and this approaches a Dirac delta function for large values of $\gamma$.

The quantitative nature of $n_0(z)$ is unknown, and we consider two scenarios to quantify $n_0(z)$ namely (a.) CER -  constant event rate  where $n_0(z)$ is independent of $z$ over the redshift range of our interest, and (b.) SFR -  where $n_0(z)\propto(0.015(1+z)^{2.7})/(1+((1+z)/2.9)^{5.6})$ traces the cosmological star formation rate \citep{madau14}.  

For every detected FRB event, we have an associated  signal to noise ratio $S/N$,  dispersion measure $DM$, pulse width $w$ and  fluence $F$.  The four telescopes which we have considered each works in a different frequency range and has a different limiting fluence for a detection, and these are all mentioned in Table~\ref{tab:1}. For each telescope, the number of detected FRB events is summarized in Table~\ref{tab:summary}, while  the observed fluence range and the $DM_{EX}$ range are  mentioned in Table~\ref{tab:1}. For convenience, we have also mentioned the $z$ range inferred  form $DM_{EX}$ assuming that $DM_{\rm host}=0$. The inferred redshifts would be smaller if a non-zero $DM_{\rm host}$ contribution were to be included.

In the present  work the  FRB population  is quantified by two  parameters $\overline{E}$ and $\gamma$, in addition to this we have the spectral index  $\alpha$ which is assumed to be the same for all the FRBs.  The aim here is to constrain these three parameters using the observed joint cumulative distribution $C_O(DM_{\rm Ex},F)$. In this paper we have used simulations to calculate the joint cumulative distribution $C_M(DM_{\rm Ex},F)$ predicted by the model.  For a    particular set of values for the   parameters $\alpha$, $\overline{E}$ and $\gamma$,we consider the null hypothesis that the observed data is drawn from the distribution predicted by the  model. We have compared  $C_M(DM_{\rm Ex},F)$ with $C_O(DM_{\rm Ex},F)$ to either  accept or reject our null hypothesis,  thereby constraining  the allowed range of the model parameters $\alpha$, $\overline{E}$ and $\gamma$. 

\subsection{Simulating the model predictions.}

The simulation considers a comoving volume extending up to $z_{\rm max}=5$  which is considerably larger than the highest redshift inferred for any of the FRB events included in our analysis (Table~\ref{tab:1}).  The angular extend of this comoving volume is equal to the full width half maxima (FWHM) of the primary beam which   is  different for each  telescope. We have populated this comoving volume with $10^6$ randomly located FRB events whose mean comoving number density follow $n_0(z)$. This provides the comoving distance $r$ and angular position $\vec{\theta}$ for each FRB. 
We have estimated the redshift $z$ from $r$ considering the $\Lambda CDM$ cosmology \citep{planck13}.
The energy $E$ associated with each event is randomly drawn from the distribution (Equation~\ref{eq:sch}).
For each FRB we estimate three observable quantities  namely (a.) the fluence $F$ from the energy $E$ using Equation~\ref{eq:fluence}, (b.) the pulse width $w$, and (c.) the extra-galactic dispersion measure $DM_{\rm Ex}$, and we use  the detection criteria (Equation~\ref{eq:Fl}) to determine whether the particular event would be detected or not.

For a FRB at redshift $z$  the observed pulse width $w=\sqrt{w_{\rm cos}^2+w_{\rm DM}^2+w_{\rm sc}^2}$ has three contributions, {\em viz.} (i) $w_i$ stretched by the cosmological expansion  $w_{\rm cos}=w_i\, (1+z)$, (ii) for observations at frequency $\nu_0$ with channel width $\Delta\nu_c$, the dispersion broadening $w_{\rm DM}=(8.3\times10^6\,{\rm DM}\,\Delta\nu_c)/\nu_0^3$, and (iii) the scatter broadening $w_{\rm sc}$.

The exact scattering mechanism in the intervening medium  is not well understood, and we consider two scattering models namely (a.) Sc-I which is based on \citet{bhat04} that  provides an empirical fit to a large number of Galactic pulsar data, here we have extrapolated this for the IGM; and (b.) Sc-II which is based on  a purely theoretical model  proposed by \citet{macquart13} considering the turbulent IGM . In both these models  $w_{\rm sc}$ dominates  $w$ at $z \ge 0.5$  (Figure 1 of \citet{bera16}), and we expect the observed pulse widths to be correlated with $z$. However such a correlation is not observed in the Parkes data \citep{cordes16}, which motivates us to also consider a third model No-Sc where there is no scattering and  $w_{\rm sc}=0$.

We next consider  $DM_{\rm Ex}=DM_{\rm IGM} + DM_{\rm Host}/(1+z)$. We have calculated $DM_{\rm IGM}=940.51\int_0^z\,\left(1+z^{\prime}\right)/ \left(\sqrt{\Omega_m\,(1+z^{\prime})^3+\Omega_{\Lambda}}\right)\,dz^{\prime}$ \citep{ioka03} assuming a completely ionized, homogeneous IGM. The constant value here will evolve with $z$ if we include He reionization at $z \approx 3$, however we do not expect this to have a very large effect on our results as all the observed FRBs are below this redshift in our analysis.

At present we have very limited information regarding the host contribution $DM_{\rm Host}$, this is also expected to vary from FRB to FRB depending on the properties of the host galaxy, the location of the FRB within the host galaxy and also the exact path traversed by the observer's line of sight.
The lowest $DM$ reported to date is $109.61\,{\rm pc\,cm}^{-3}$ for the FRB $180729$ detected at CHIME. This indicates  that $DM_{\rm Host}$ may not exceeds the value $100\,{\rm pc\,cm}^{-3}$ \citep{macquart20}. In this work we consider two scenarios for $DM_{\rm Host}$ namely (a.) DM50 where all the FRBs have fixed $DM_{\rm Host}=50\,{\rm pc\,cm}^{-3}$; and (b.) DMrand where $DM_{\rm Host}$ values are randomly  drawn 
from a Gaussian distribution with mean $\overline{DM}_{\rm Host}= 50\,{\rm pc\,cm}^{-3}$ and root mean square value $\Delta {DM}_{\rm Host}= 15 \,{\rm pc\,cm}^{-3}$. The distribution is cut off at $0$ and $100\,{\rm pc\,cm}^{-3}$ at the two ends respectively.

\begin{figure*}
\centering
\includegraphics[scale=0.7]{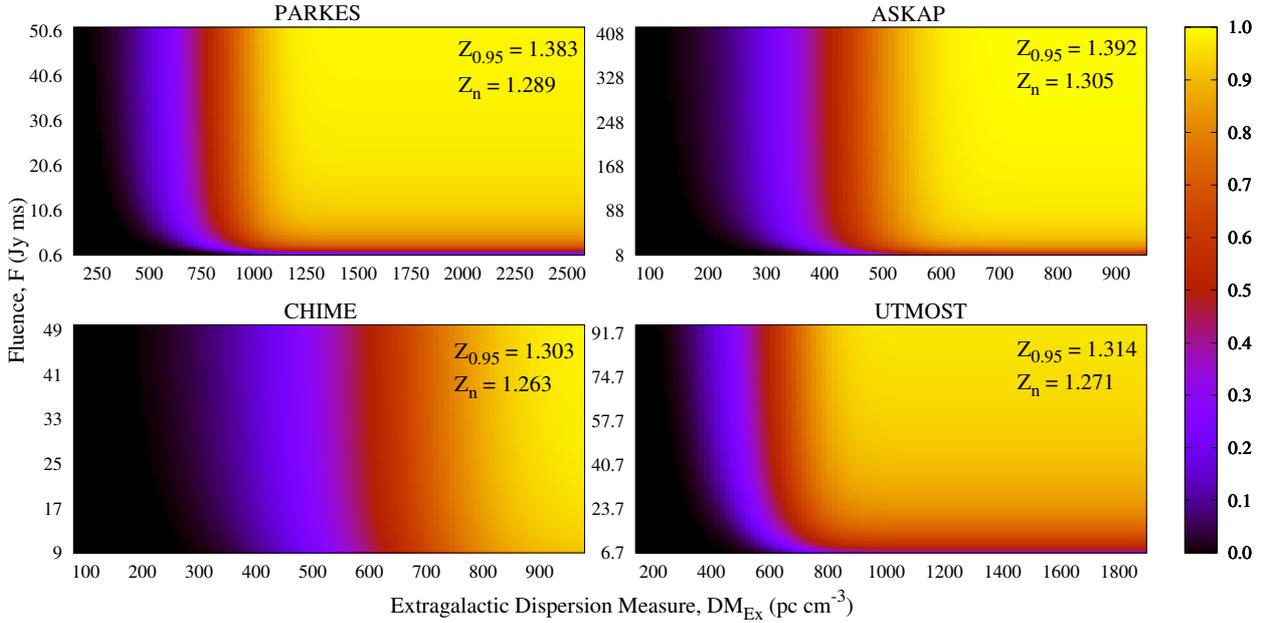}
\caption{The model prediction for the  joint cumulative distribution $C_{M}(DM_{\rm Ex},F)$ for FRB detection at Parkes, ASKAP, CHIME and UTMOST. For this prediction we consider the model parameter values  $\overline{E}=10^{33}\,{\rm J}$, $\alpha=-2$ and $\gamma=5$ with CER, DM50 and No-Sc.  The values of $Z_n$ and $Z_{0.95}$ are shown in the respective panels, we see that this particular model is consistent with the observations $(Z_n < Z_{0.95})$  for all the four telescopes.}
\label{fig:3.1}
\end{figure*}

Figure~\ref{fig:3.1} shows $C_{M}(DM_{\rm Ex},F)$ for the model parameters $\alpha=-2$, $\E33=1$, and $\gamma=5$ with CER, DM50 and No-Sc. Note that throughout we shall use $E_{33}$ which is in units of $10^{33} \, J$. As mentioned in Figure~\ref{fig:3.1}. 
the different panels show $C_{M}(DM_{\rm Ex},F)$ for the four different telescopes considered in this paper. We  may visually compare 
Figures~\ref{fig:3.1} and \ref{fig:2.1}, however it is not obvious if the model prediction match the observations  and whether we may accept or reject this particular set of parameter values.

\begin{figure}
    \centering
    \includegraphics[width=\columnwidth]{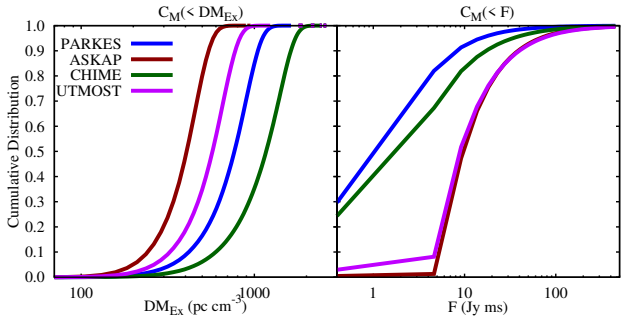}
    \caption{The model predictions for the  cumulative distribution of the extragalactic dispersion measure $C_M(<DM_{\rm Ex})$ (left) and fluence $C_M(<F)$ (right) for  FRB detection  at Parkes (blue), ASKAP (red), CHIME (green) and UTMOST (magenta). For this prediction we consider the model parameter values $\E33=1$, $\alpha=-2$ and $\gamma=5$ with CER, DM50 and No-Sc.}
    \label{fig:3.2}
\end{figure}

Figure~\ref{fig:3.2} shows the 1D cumulative distributions $C_{M}(<DM_{\rm Ex})$ and $C_{M}(<F)$ for the same values of the model parameters as Figure~\ref{fig:3.1}. The  different lines correspond to different telescopes as mentioned in the figure. Comparing  $C_{M}(<DM_{\rm Ex})$ with the observations (Figure \ref{fig:2.2}) we find that they are qualitatively similar for  Parkes, ASKAP, and UTMOST. For CHIME the $DM_{\rm Ex}$ values predicted by the model are larger compared to the observed values. Considering $C_{M}(<F)$  we find that the model predictions are in qualitative agreement with the observations (Figure \ref{fig:2.2})  for  Parkes  and UTMOST. For both ASKAP and CHIME the 
$F$ values predicted by the model are smaller  compared to the observed values. Although this graphical representation visually demonstrates the difference between the model predictions and the observation, a more quantitative analysis is required to rule out or accept the null hypothesis.

\subsection{Two dimensional KS test.}
\begin{figure}
\centering
\includegraphics[width=\columnwidth]{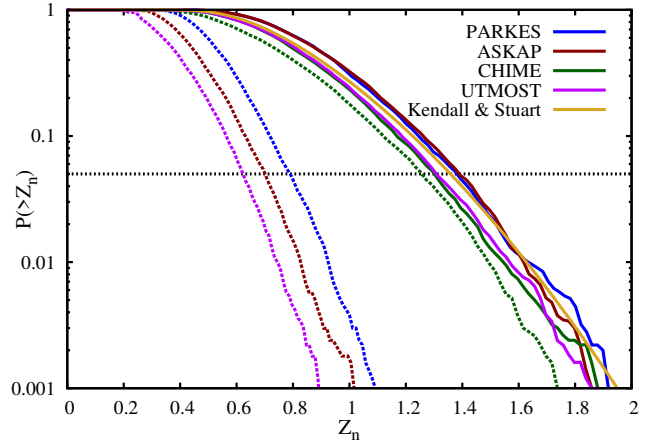}
\caption{The statistic of $Z_n$, $P(>Z_n)$, for the FRBs assume to detected at Parkes (blue), ASKAP (red), CHIME (green) and UTMOST (magenta). For this prediction we consider two sets of model parameter values (A.) $\alpha=-2$, $\E33=1$, $\gamma=5$ and (B.) $\alpha=10$, $\E33=100$, $\gamma=5$ with CER, DM50 and No-Sc. The solid and dashed color line denote the set (A.) and the set (B.) respectively. The black dashed line corresponds to $P(>Z_n)=0.05$.} 
\label{fig:3.3}
\end{figure}

We use the two dimensional (2D) Kolmogorov-Smirnov (KS) test to test the null hypothesis that the observed data is drawn from the distribution predicted by our model. In comparison to the 2D KS test the one dimensional (1D) KS test is relatively straight forward \citep{conover99}. In the 1D KS test we consider the observed and predicted 1D cumulative distributions $C_{O}(<x)$ and $C_{M}(<x)$ respectively. The test statistic $Z_n$ of the 1D KS test is the maximum difference between $C_{O}(<x)$ and $C_{M}(<x)$ scaled  with $\sqrt{n}$ , \textit{i.e.} $Z_n=\sqrt{n} \times D_n$ where $D_n = {\rm max} |C_{O}(<x)-C_{M}(<x)|$ and $n$ is the number of data points. Considering the 1D KS test, $P(>z)$  
which  is the probability that $Z_n$ exceeds $z$ is predicted to have an 
asymptotic form \citep{kendall46}  given by
\begin{equation}
    P_{\rm as}(>z)=2\sum_{k=1}^{\infty}(-1)^{k-1}\exp\left({-2\,k^2 z^2}\right)\,. 
    \label{eq:Zn}  
\end{equation}
This is shown by the yellow line in  Figure \ref{fig:3.3}. In order to test the null hypothesis at a significance level $\beta$, we first determine $z_{\beta}$ for which $P(>z_{\beta})=1-\beta$. One rejects the null hypothesis with a significance level $\beta$ if $Z_n$ exceeds the  value $z_{\beta}$. For example, the null hypothesis will be rejected with $95\%$ confidence if $Z_{n}>z_{0.95}=1.3$. 

The 2D KS test is  relatively more involved in comparison to the 1D KS test.  We consider a set of $n$  observed data points   $(x_i,y_i)$ with  $i=1,2,...,n$.  Following the procedure discussed in \citet{peacock83}, we  use the test statistic $Z_n=\sqrt{n}\,D_n$ where  $D_n$ is the maximum difference between the observed and predicted joint cumulative distributions. We now discuss how we have calculated $D_n$. 
Considering any  data point $(x_i,y_i)$, we separately compare the cumulative distributions in  four quadrants centred around this data point. For example,  in the  second (II) quadrant $(x \le x_i,y>y_i)$ we define the observed  cumulative distribution $C_{O}^{II}(x_i,y_i)$ which is the fraction of data points with $x \le x_i$ and $y > y_i$, and this is compared with the corresponding model prediction. A similar comparison is carried out for all the quadrants for all the data points. $D_n$ refers to the absolute value of the maximum difference between the observed and model cumulative distributions across all the quadrant and all the data points.  

The statistics of $Z_n$  is not universal for the 2D KS test, and we use simulations to estimate the integral probability $P(>Z_n)$. For any given telescope, our simulations give  a  total sample of detectable FRB events roughly $5\times10^5$  in number.    We use this total sample to create a mock detected sample which contains the number of FRBs actually detected  at the  given telescope. The mock detected samples are created by randomly drawing a subset of total sample of detectable FRB . For any given telescope, we have generated  $5000$ mock detected samples which were used to calculate the statistics of $Z_n$.   Figure~\ref{fig:3.2} shows $P(>Z_n)$ for the all the four telescopes considered here for  two sets of model parameters values (A.) $\alpha=-2$, $\E33=1$, $\gamma=5$ and (B.) $\alpha=10$, $\E33=100$, $\gamma=5$. We have considered  CER with DM50 and No-Sc for  both  (A.) and (B.).
For (A.) we see that  $P(>Z_n)$ is very similar for all the four telescopes considered here. Further, we also note that $P(>Z_n)$ is very close to $P_{\rm as}(>z)$ which is  predicted for the 1D KS test (eq. \ref{eq:Zn}). However, for (B.) we see that the predicted $P(>Z_n)$ is  close to $P_{\rm as}(>Z_n)$ only for CHIME whereas the  $P(>Z_n)$  predicted for the three other telescopes fall off much faster and  are quite different   from $P_{\rm as}(>Z_n)$.  This illustrates why it is necessary to use simulations to  estimate the $Z_n$ statistics individually for each set of parameter values.  We note that our subsequent  analysis shows that model (A) is consistent with the observations for all four telescopes, while  model (B) is only consistent with CHIME and it is ruled out by the observations at the three other telescopes.

We now define the  quantity $Z_{0.95}$ such that  $P(>Z_{0.95})=0.05$, {\it i.e.} the null hypothesis is ruled out at $95 \%$ confidence if $Z_n \ge Z_{0.95}$.  Considering Figure~\ref{fig:3.3} we see that  the intersection of  the  $P(>Z_n)$ curves with the  horizontal black dashed line  gives the value of $Z_{0.95}$. For (A.) we find that the values of $Z_{0.95}$ are very close to  each other for the four telescopes considered here, and we find $Z_{0.95}= 1.382, 1.389, 1.303$ and $1.308$ for Parkes, ASKAP, CHIME, and UTMOST  respectively while $Z_{0.95}=1.362$ for the 1D KS test. In contrast for (B.) we find  the values $Z_n=0.792,0.702,1.288$ and $0.624$ for the four respective telescopes. 

In the subsequent analysis,  considering  a set of model parameter values for each telescope  we have compared the observed FRB sample with the total simulated sample of detectable FRBs  ($\sim 5 \times 10^5$ in number) predicted for the model and have used this  to calculate $Z_n$. The statistics of $Z_n$ and the value of $Z_{0.95}$ is estimated using $5,000$  simulated sub-samples each containing the same number of FRBs as the observed FRB sample. The particular set of model parameter values is  ruled out at $95 \%$ confidence if $Z_n \ge Z_{0.95}$, else the model is accepted.

\section{Results}\label{sec:4}
Here for each telescope we identify  the allowed region of  $(\alpha, \E33)$ space considering different values of $\gamma$. The null hypothesis that the observed FRB data is  drawn from the distribution predicted by our model is accepted within this region, and the exterior region is ruled out at $95\%$ confidence. We focus on the common region in  $(\alpha, \E33)$ space where the null hypothesis is simultaneously satisfied for the four different telescopes considered here. For this analysis we consider values of $\alpha$ in the range $-10\leq\alpha\leq 10$ and  $\E33$ in the range $10^{-2}\leq\E33\leq10^{2}$. We have considered $\gamma$ in the range $0\leq\gamma\leq10$, however we have  shown the results only for a few representative values  $\gamma=0,3,5$ and $10$ which span the entire range. 

\begin{figure}
    \centering
    \psfrag{Mean Energy, yy}{Mean Energy, $\E33$}
    \psfrag{Spectral Index, xx}{Spectral Index, $\alpha$}
    \includegraphics[width=\columnwidth]{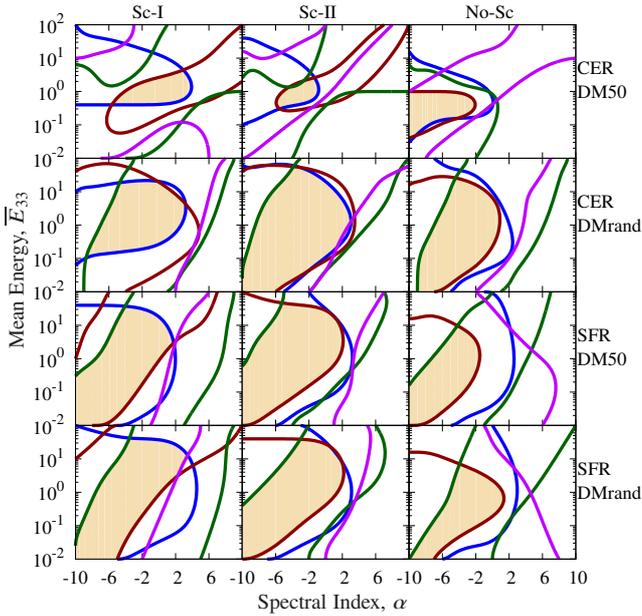}
    \caption{The solid contour enclose region of the $(\alpha,\E33)$ space where the null hypothesis is accepted and the exterior region is ruled out at $95\%$ confidence for Parkes (blue), ASKAP (red), CHIME (green) and UTMOST (magenta). Different panels show the result for different combinations of FRB rate and DM model considering $\gamma=0$.}
    \label{fig:4.0}
\end{figure}

Figure~\ref{fig:4.0} shows the contours of the allowed region of $(\alpha,\E33)$ space for the four different telescopes considered here. Each panel of the figure corresponds to the different combinations of FRB rate, DM and scattering models as indicated in the figure. 
All the panels in this figure correspond to a fixed value  $\gamma=0$, the figures in Appendix \ref{appendix1}  show the $\gamma$  dependence of the results. The first row of the figure shows the results  for CER  DM50, we  first consider the top left panel which corresponds to Sc-I.
We see that for Parkes (blue) the allowed  $(\alpha,\E33)$ values are bounded within $\alpha \le 4$ and  $\E33 \ge 0.4$,  $\E33$  also has an upper bound which increases as $\alpha$ is decreased. 
Considering  Sc-II and No-Sc, we see that the allowed region is similar to that for Sc-I,  however the   allowed $\alpha$ range is reduced  and in most cases $\alpha>0$ is ruled out. Considering the results for ASKAP (red) we see that the  allowed $\alpha$ range is bounded from the left ($\geq-6$) for Sc-I and Sc-II, whereas it is bounded from the right  $(\le -2)$ for No-Sc. In all cases the $\E33$ values  are bounded within a narrow range which has some overlap with the region allowed by Parkes. 
Considering CHIME (green)  we see that the allowed ($\alpha,\E33$) region is a nearly diagonal band  for Sc-I and Sc-II, whereas  it  is bounded from the right and the top for No-Sc.  We see that  the allowed region for CHIME has some overlap with common region of Parkes and ASKAP. 
Considering  UTMOST (magenta) we see that for Sc-I, Sc-II and No-Sc the allowed region is a broad nearly diagonal band which  completely encompasses the common allowed region  for Parkes, ASKAP and CHIME and does not provide any additional constraint. We see that for Sc-I and Sc-II the common allowed region for all the four telescopes is bounded with the $(\alpha,\E33)$ range considered here. However, for No-Sc this region is not bounded from the left  {\it i.e.} we do not have a lower limit on $\alpha$ within  the $(\alpha,\E33)$ range considered here.

The second, third and fourth rows of Figure~\ref{fig:4.0}  show the results for CER DMrand, SFR DM50 and SFR DMrand respectively. Considering these together, we find that  typically the $\alpha$ range is bounded from the right by  both Parkes (blue) and ASKAP (red), and in many cases these two telescopes also impose upper and lower bounds on the allowed $\E33$ range.   The allowed region for CHIME (green) typically is a diagonal band which runs across the $(\alpha,\E33)$ plane, and in most cases this bounds the common allowed region from the left.  The allowed region for UTMOST (magenta) is rather broad, and in most cases this does not impose any additional constraint on the common allowed region for the other three  telescopes.

We see that we have the tightest constraints in the top row which shows CER DM50 for which  the comoving event rate density does not evolve with redshift, and $DM_{\rm Host}$ is the same for all the FRBs.  Considering CER and DMrand (second row of Figure~\ref{fig:4.0}), this model allows for a spread in  $DM_{\rm Host}$ and this is reflected in a broadening of the allowed parameter range. In all cases, CHIME bounds the common allowed region from the left whereas the bounds in the other directions come primarily from Parkes and ASKAP.  For Sc-I the common allowed region is completely bounded within the parameter range shown here, however for Sc-II and No-Sc this region  is not completely bounded within this  range.  For Sc-II and No-Sc,  the allowed parameter range extends beyond the bottom left corner of  the figure  $(\alpha <-6, \E33 \le 3 \times 10^{-2})$.

 We next consider SFR DM50 (third row of Figure~\ref{fig:4.0}) for which the comoving event rate density traces  the observed redshift evolution of the SFR, and $DM_{\rm Host}$ is the same for all the FRBs. We see that the predictions for the individual telescopes are different from those in the two earlier rows. Considering the common allowed region, for Sc-I this is larger than that predicted for CER DMrand, whereas for Sc-II and No-Sc this is comparable. In all cases the common allowed range is bounded from the top and the left, however it extends beyond the bottom left corner of the figure. Considering SFR and DMrand (forth row of Figure~\ref{fig:4.0}), we see that the allowed $(\alpha,\E33)$ regions for the individual telescopes, and also the common allowed region,  are very similar to  those for SFR  DM50.
 
\begin{figure}
    \centering
    \psfrag{Mean Energy, yy}{Mean Energy, $\E33$}
    \psfrag{Spectral Index, xx}{Spectral Index, $\alpha$}  
    \includegraphics[width=\columnwidth]{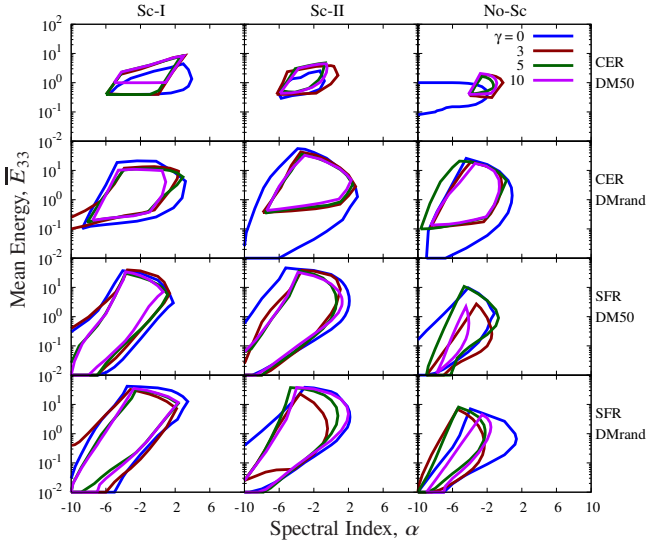}
    \caption{The contours of the common allowed $(\alpha,\E33)$ regions where the null hypothesis is simultaneously satisfied for the four telescopes considered in this paper. Different panels correspond to different combinations of FRB rates, $\gamma$, DM and scattering models mentioned in the figure.}
    \label{fig:4.5}
\end{figure}

The entire discussion till now has been restricted to a single value $\gamma=0$. The figures in Appendix \ref{appendix1} shows the individual telescope predictions, and also the common allowed region, for $\gamma=0,3,5$ and $10$.
Considering all the different cases considered here, the common allowed   $(\alpha,\E33)$ regions are shown in Figure \ref{fig:4.5}. The intrinsic FRB energy distribution (eq.~\ref{eq:sch}) is broadest for $\gamma=0$, and it gets narrow as $\gamma$ is increased. We see that this is reflected in the allowed $(\alpha,\E33)$ regions which are largest for $\gamma=0$ and typically get smaller as $\gamma$ is increased. Table~\ref{tab:2} presents the bounding $\alpha$ and $\E33$ values which enclose the allowed regions for $\gamma=0$ and $\gamma=10$. Note that the allowed regions are not rectangles, and they only cover a part of the rectangles  enclosed by the values in Table~\ref{tab:2}.  We see that we have the tightest constraints if we consider CER with DM50 for which the $(\alpha,\E33)$ regions are smallest for No-Sc where  $-4.2 \le  \alpha \le -0.9$ and $0.4 \le \E33 \le 2.1$ for $\gamma=10$. The allowed $(\alpha,\E33)$ range  increase substantially, particularly towards smaller $\alpha$ if we consider the other scattering models.
For CER the allowed region also increases substantially if we  allow a spread in $DM_{\rm Host}$. Considering SFR with DM50 we see that the allowed regions are substantially larger than CER with DM50 and they are comparable to CER with DMrand. Unlike CER, for SFR the allowed region does not  increase very significantly  if we allow a spread in $DM_{\rm Host}$. 
Considering the lower  three rows of the figure, we see that the extent of the allowed regions increases substantially in comparison to CER with DM50. This increase is predominantly along a diagonal extending towards the bottom left corner {\it ie.}  negative $\alpha$ and lower $\E33$ values.  Considering all the different cases considered here,  in every case the common allowed region includes the range  $-3.9\leq\alpha\leq-1.3$ and $0.42\leq\E33\leq1$.  Further, in all cases large values $\alpha > 4$ and $\E33 >60$ are ruled out. 

\begin{figure}
    \centering
    \psfrag{Mean Energy, yy}{Mean Energy, $\E33$}
    \psfrag{Spectral Index, xx}{Spectral Index, $\alpha$}
    \includegraphics[width=\columnwidth]{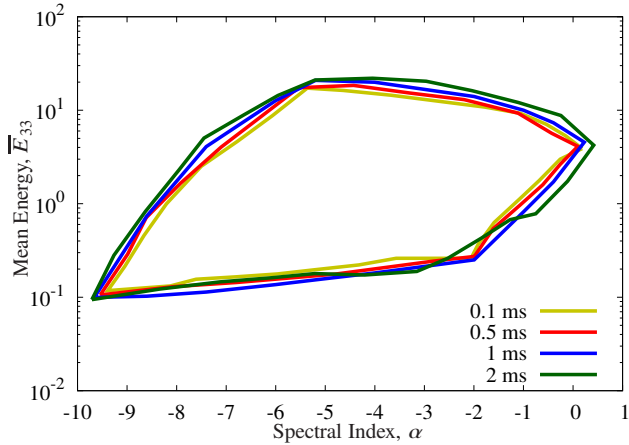}
    \caption{The  common allowed region  for different values of the FRB intrinsic pulse width $w_i$ 
    mentioned in the figure. We consider CER and DMrand with $\gamma=5$ and No-Sc for this comparison.}
    \label{fig:5.1}
\end{figure} 
 
The entire discussion till now has been restricted to the situation where the intrinsic pulse width has been fixed at $w_i=1\, {\rm ms}$. We have also repeated the same analysis for $w_i=0.1$ ms, $0.5$ ms and $2$ ms for which Figure \ref{fig:5.1} shows the results for one case where we have used CER and DMrand with $\gamma=5$ and No-Sc. We see that the common allowed $(\alpha,\E33)$ region changes by only a small amount if $w_i$  is varied, and we have not shown this explicitly for all the other cases considered here.  

\begin{figure}
    \centering
    \psfrag{Mean Energy, yy}{Mean Energy, $\E33$}
    \psfrag{Spectral Index, xx}{Spectral Index, $\alpha$}
    \includegraphics[width=\columnwidth]{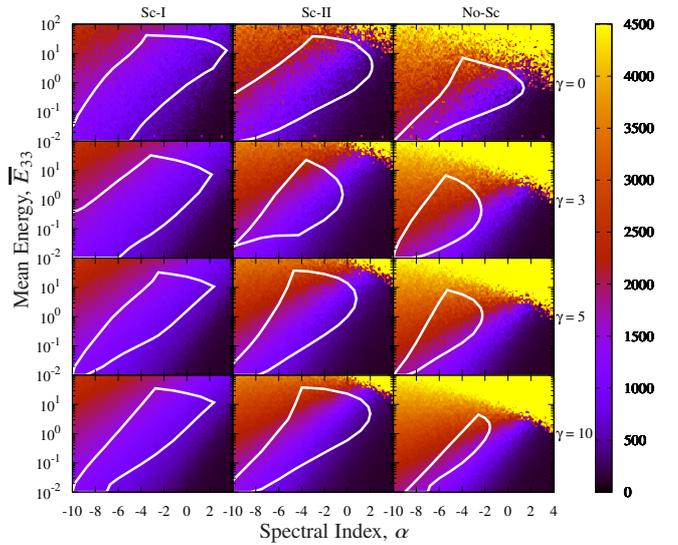}
    \caption{Considering CHIME, this shows $[DM_{Ex}]_{\rm max}$ as a function of $(\alpha,\E33)$ for different scattering models and values of $\gamma$  as indicated for each panel. For this we have assumed  $w_i=1\,{\rm ms}$ and $DM_{\rm Host}=100 \,\,{\rm pc\,cm}^{-3}$.}
    \label{fig:5.2}
\end{figure} 

Considering the large number of detected FRBs anticipated  to be  reported from CHIME, we investigate if our models make any unique prediction which can be used to validate or falsify the results of the present analysis.  Here we focus on  $[DM_{Ex}]_{\rm max}$ which is the maximum value of $DM_{Ex}$ predicted for the FRBs detected by CHIME. For this, we have used $w_i=1\,{\rm ms}$ and assumed that the maximum possible  host contribution to be  $DM_{\rm Host}=100 \,\,{\rm pc\,cm}^{-3}$. Figure~\ref{fig:5.2} shows the variation of $[DM_{Ex}]_{\rm max}$ with  $(\alpha,\E33)$ for different combinations of $\gamma$ and scattering model. Note that the value of $[DM_{Ex}]_{\rm max}$ does not change depending on whether we use CER or SFR for the event rate density. We see that $[DM_{Ex}]_{\rm max}$ increases with $\E33$ {\it i.e.} the FRBs extend out to a higher redshift, further in most cases $[DM_{Ex}]_{\rm max}$ increases if $\alpha$ is reduced.  Considering all the combinations of $\gamma$ and scattering models considered here, we find that $[DM_{Ex}]_{\rm max}$ does not exceed $3700\,{\rm pc\,cm}^{-3}$    within the common allowed region (white solid line) of $(\alpha,\E33)$ space.  The models outside the common allowed region have been ruled out at $95 \%$ confidence, and it is extremely unlikely that  the  $DM_{Ex}$ measured for the FRBs at CHIME exceeds $3700\,{\rm pc\,cm}^{-3}$. The detection of  FRBs  with larger  $[DM_{Ex}]_{\rm max}$ at CHIME would bring into question the assumptions of the present analysis.

\begin{table*}
    \caption{The allowed range of $\alpha$ and $\E33$ for which the null hypothesis is simultaneously satisfied for Parkes, ASKAP, CHIME and UTMOST considering $\gamma=0\,{\rm and}\,10$ with different combinations of FRB rates, DM and scattering models mentioned in the table.}
    \centering
    \begin{tabular}{cccccccccccccccccccc}
    \hline
    FRB & DM & $\gamma$ & \multicolumn{8}{c}{Spectral Index, $\alpha$} & & \multicolumn{8}{c}{Mean Energy, $\overline{E}_{33}$} \\
     Rate & Model &  & \multicolumn{2}{c}{Sc-I} & & \multicolumn{2}{c}{Sc-II} & & \multicolumn{2}{c}{No-Sc} & & \multicolumn{2}{c}{Sc-I} & & \multicolumn{2}{c}{Sc-II} & & \multicolumn{2}{c}{No-Sc} \\
     & & & min & max & & min & max & & min & max & & min & max & & min & max & & min & max \\ 
     \hline
     & DM50 & $0$ & $-5.6$ & $3.9$ & & $-6.1$ & $-0.8$ & & - & $-2.0$ & & $0.4$ & $4.4$ & & $0.3$ & $2.6$ & & - & $1.0$ \\
     & & $10$ & $-5.1$ & $3.0$ & & $-5.9$ & $-0.5$ & & $-4.2$ & $-0.9$ & & $1.0$ & $8.3$ & & $0.4$ & $4.4$ & & $0.4$ & $2.1$ \\ 
    CER & & & & & & & & & & & & & & & & & & & \\ 
     & DMrand & $0$ & $-8.7$ & $3.2$ & & - & $3.1$ & & $-9.0$ & $0.9$ & & $0.1$ & $21.3$ & & - & $56.2$ & & - & $25.9$ \\
     & & $10$ & $-7.5$ & $0.9$ & & $-7.5$ & $2.3$ & & $-8.5$ & $-0.6$ & & $0.2$ & $11.0$ & & $0.4$ & $33.1$ & & $0.1$ & $17.5$ \\     
     \hline
     & DM50 & $0$ & - & $1.8$ & & - & $2.1$ & & - & $-1.2$ & & - & $37.4$ & & - & $46.6$ & & - & $9.8$ \\
     & & $10$ & - & $0.6$ & & - & $1.3$ & & $-7.8$ & $-4.1$ & & - & $34.2$ & & - &  $33.1$ & & - & $2.2$ \\     
    SFR & & & & & & & & & & & & & & & & & & & \\ 
     & DMrand & $0$ & - & $3.5$ & & - & $2.2$ & & - & $1.4$ & & - & $42.1$ & & - & $38.6$ & & - & $7.1$ \\
     & & $10$ & - & $2.4$ & & - & $2.2$ & & - & $-1.6$ & & - & $35.4$ & & - & $38.3$ & & - & $4.5$ \\    
     \hline
    \end{tabular}
    \label{tab:2}
\end{table*}

\section{Summary and Conclusion}\label{sec:5}
The various telescopes which have detected FRBs each works at a different frequency band with a  different field of view and limiting fluence for a detection.   We  present a methodology to combine the constraints on  the properties of the FRB population arising from 
observations at  different telescopes. In this paper  we have used the FRBs detected at Parkes, ASKAP, CHIME and UTMOST  to constrain 
the allowed values of the spectral index $\alpha$ and mean FRB energy $\E33$ considering different values of $\gamma$  which is   the slope of the  Schecter luminosity function for the FRB energies. 
 Further, we consider three different scenarios for the pulse broadening due to scattering during propagation (Sc-I, Sc-II and No-Sc),  two possible scenarios for the event rate density variation with $z$  (CER and SFR) and two scenarios for $DM_{\rm Host}$ (DM50  and DMRand). For each telescope we  have quantified the observed  FRB distribution  using the 2D joint cumulative distribution $C_{O}(DM_{\rm Ex},F)$ for the measured extragalctic dispersion measure $DM_{\rm Ex}$ and fluence $F$. The simulated 
  model predictions were similarly quantified using  $C_{M}(DM_{\rm Ex},F)$. We have used a 
 two dimensional Kolmogorov-Smirnov test for the  null hypothesis  that  the observed date is drawn from the distribution predicted by our model. This was used to rule out regions of the $(\alpha,\E33)$ space at $95 \, \%$ confidence for each telescope, and finally identify the combined allowed region for all four telescopes. The entire analysis was carried out assuming the same  intrinsic pulse width $w_i=1 \, {\rm ms}$  for all the FRBs,  we have  checked that the results do not change much if we use $w_i=0.1$ ms,  $0.5$ ms or $2$ ms instead.

 We find the constraints from UTMOST are rather weak, and these do not impose any  restriction in addition to those  imposed by Parkes, ASKAP and CHIME combined.  In all cases we find that there is a combined allowed region of 
 $(\alpha,\E33)$ space which is consistent with the FRBs observed  at all the four telescopes.   The $(\alpha,\E33)$  values outside the common allowed region identified by our analysis are ruled out at $95 \%$ confidence.  
 The constraints are tightest for CER 
 with DM50, No-Sc  and  $\gamma=5$  where the allowed region is bounded within  $-4 \le \alpha \le -1.8$ and $0.3 \le \E33 \le 1$.
 The allowed region increases if $\gamma$ is reduced, the FRB event rate density is changed,  $DM_{\rm Host}$ has a spread  or if   we consider a different scattering model. Altogether  the allowed $(\alpha,\E33)$ range  predominantly increases  along a diagonal extending towards the bottom left corner {\it ie.}  negative $\alpha$ and lower $\E33$ values. In many cases the common allowed region is not bounded to the left and/or bottom with the range of $(\alpha,\E33)$ values considered here. 
For  all the different cases considered here,  in every case the common allowed region includes the range  $-3.9\leq\alpha\leq-1.3$ and $0.42\leq\E33\leq1$.  In all cases large values $\alpha > 4$ and $\E33 >60$ are ruled out,  and  it is extremely unlikely that the FRB populations has spectral indices  $\alpha > 4$ and  energies $\E33 >60$. 

Considering the large number of detected FRBs anticipated  to be  reported from CHIME, we predict that the $DM_{Ex}$ measured for the FRBs at CHIME is extremely unlikely to exceed $3700\,{\rm pc\,cm}^{-3}$. The detection of  FRBs  with a larger $[DM_{Ex}]_{\rm max}$ at CHIME would bring into question the assumptions of the present analysis. 
With the advent of more FRB data we expect this method to yield tighter constraints on the intrinsic properties of the FRB population, the scatter broadening of the pulse as it propagates and also the distribution of $DM_{\rm Host}$.


\appendix
\section{$\gamma$ dependence of the results}
\label{appendix1}
\begin{figure}
    \centering
    \psfrag{Mean Energy, yy}{Mean Energy, $\E33$}
    \psfrag{Spectral Index, xx}{Spectral Index, $\alpha$}
    \includegraphics[width=\columnwidth]{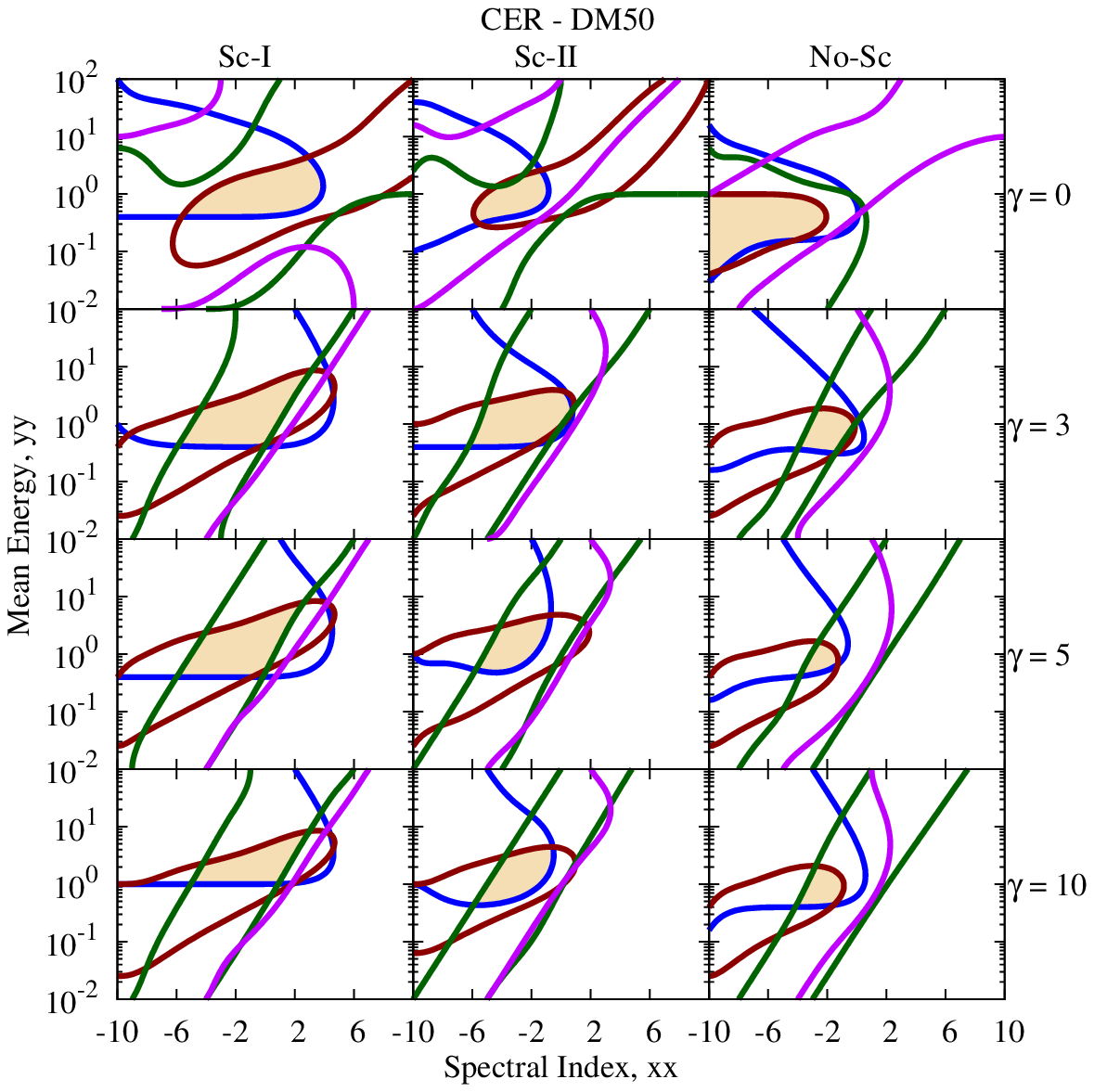}
    \caption{The solid contour enclose region of the $(\alpha,\E33)$ space where the null hypothesis is accepted and the exterior region is ruled out at $95\%$ confidence for Parkes (blue), ASKAP (red), CHIME (green) and UTMOST (magenta). Different panels show the result for different combinations of $\gamma$ and scattering model with CER and DM50.}
    \label{apnx:1}
\end{figure}

\begin{figure}
    \centering
    \psfrag{Mean Energy, yy}{Mean Energy, $\E33$}
    \psfrag{Spectral Index, xx}{Spectral Index, $\alpha$}
    \includegraphics[width=\columnwidth]{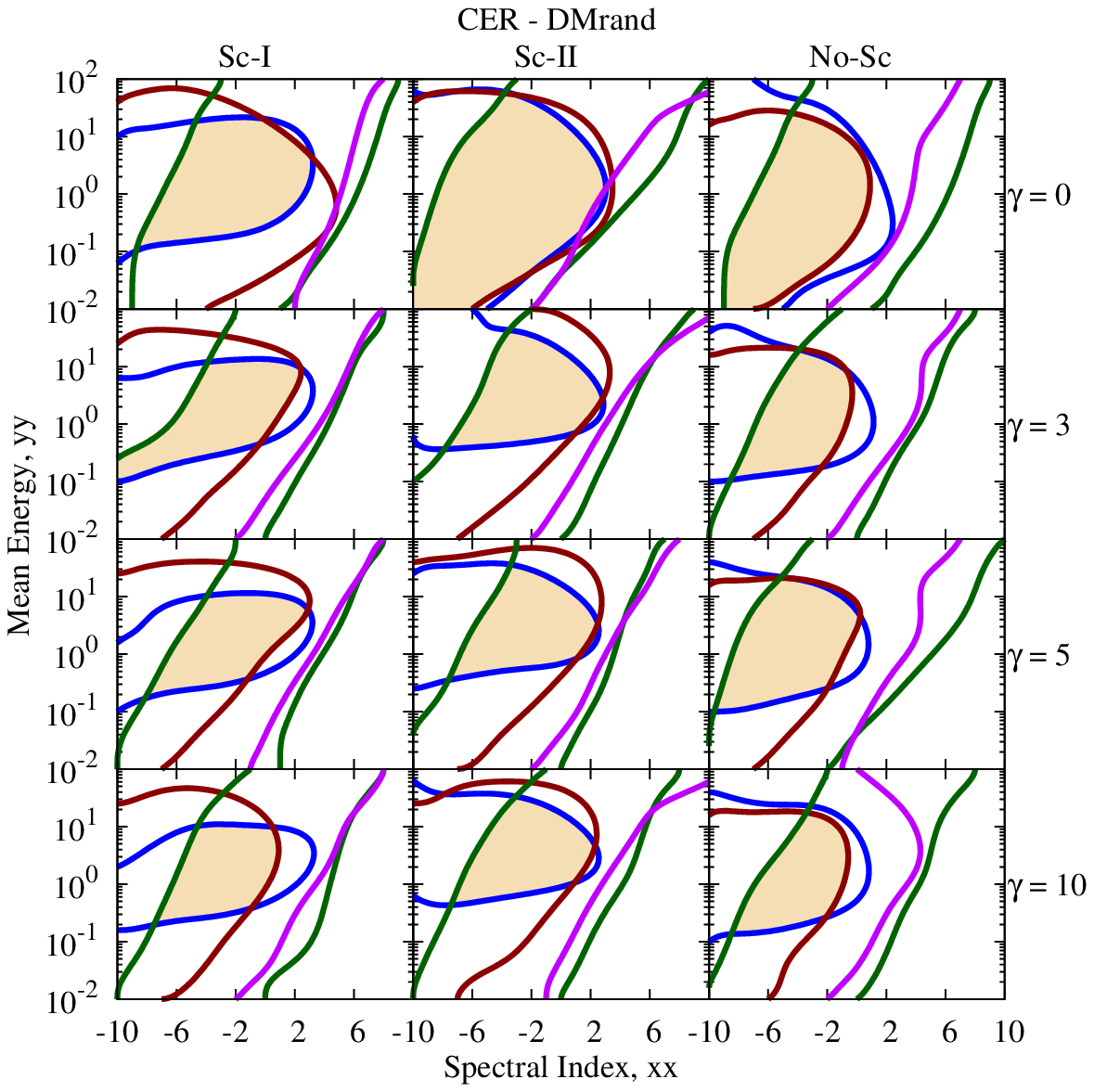}
    \caption{Same as Figure~\ref{apnx:1} for CER and DMrand.}
    \label{apnx:2}
\end{figure}

\begin{figure}
    \centering
    \psfrag{Mean Energy, yy}{Mean Energy, $\E33$}
    \psfrag{Spectral Index, xx}{Spectral Index, $\alpha$}    
    \includegraphics[width=\columnwidth]{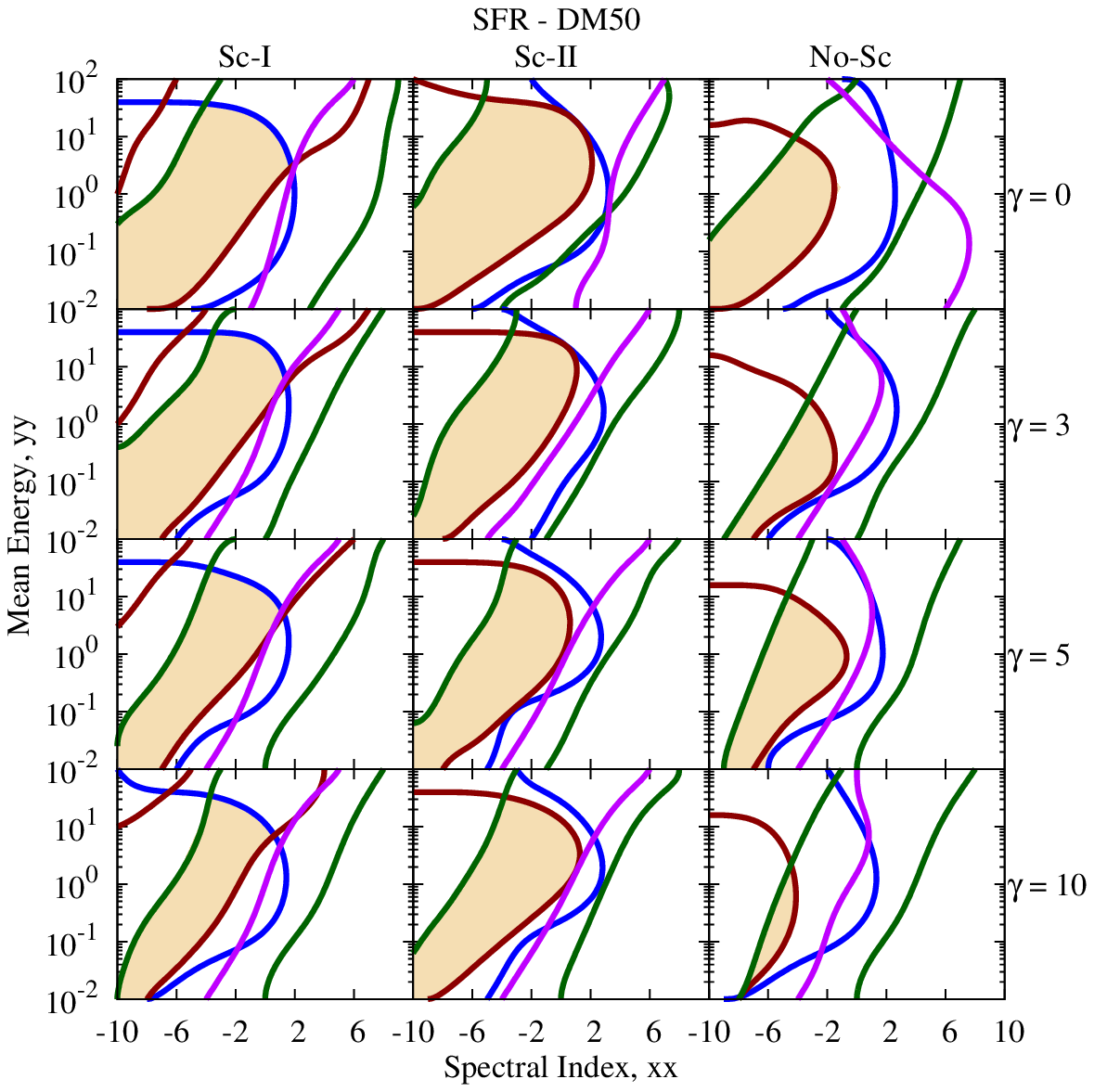}
    \caption{Same as Figure~\ref{apnx:1} for SFR and DM50.}
    \label{apnx:3}
\end{figure}

\begin{figure}
    \centering
    \psfrag{Mean Energy, yy}{Mean Energy, $\E33$}
    \psfrag{Spectral Index, xx}{Spectral Index, $\alpha$}    
    \includegraphics[width=\columnwidth]{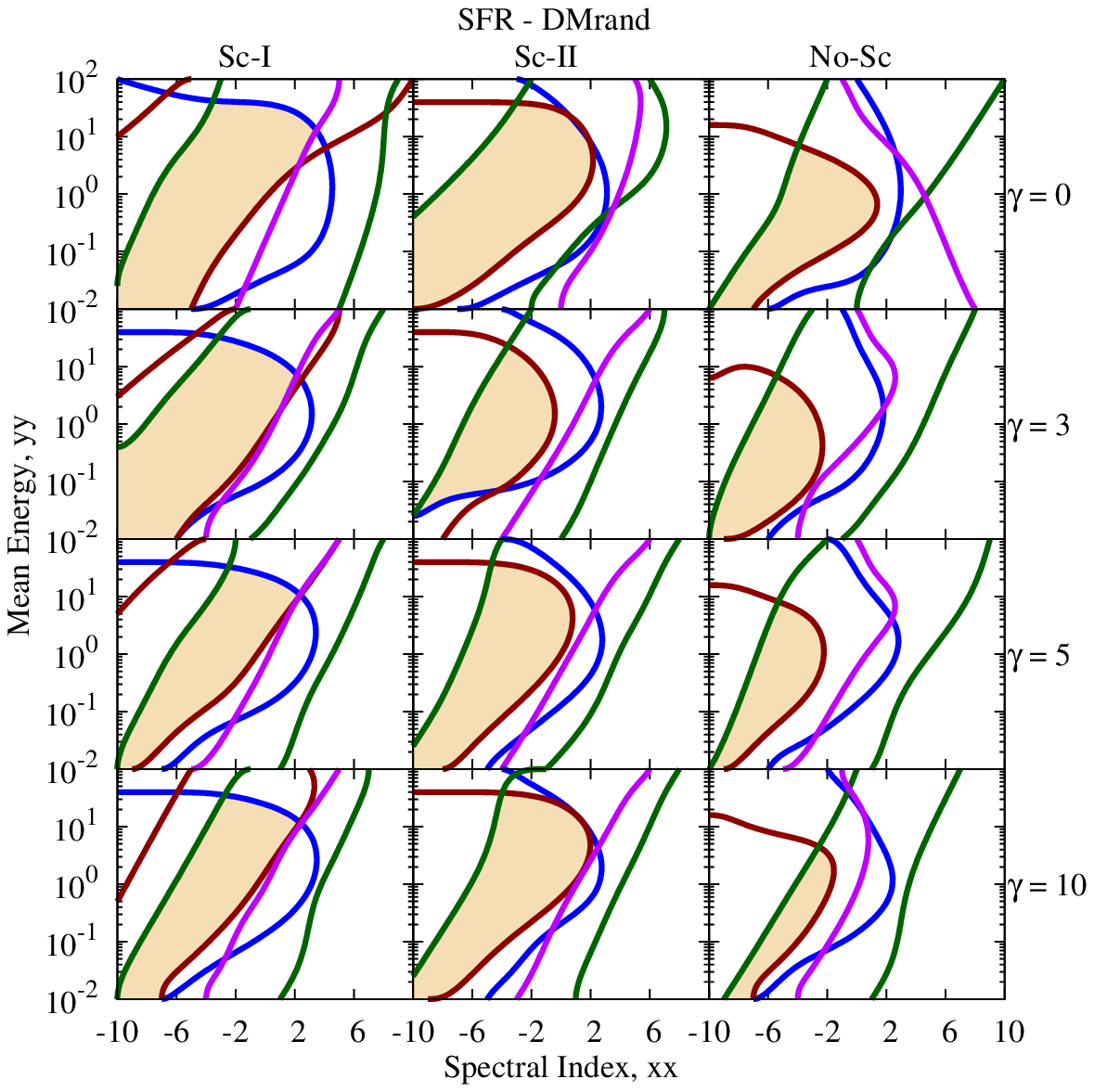}
    \caption{Same as Figure~\ref{apnx:1} for SFR and DMrand.}
    \label{apnx:4}
\end{figure}

Considering each instrument Parkes, ASKAP, CHIME and UTMOST,  we show   how the allowed region of $(\alpha,\E33)$ space varies with  $\gamma$ for values in the range $0\leq\gamma\leq 10$. The results are shown for only a few representative values $\gamma=0,3,5$ and $10$ which span the entire range. Figures \ref{apnx:1}, \ref{apnx:2}, \ref{apnx:3} and \ref{apnx:4}  respectively correspond to  CER with DM50, CER with DMrand, SFR with DM50 and SFR with DMrand. Considering any particular  figure,  each panel corresponds to different values of $\gamma$ and a different scattering model,  as indicated in the figures.

\label{lastpage}


\begin{thebibliography}{mnras}
\bibitem[\protect\citeauthoryear{Agarwal et al.}{2019}]{agarwal19} 
Agarwal, D., Lorimer, D. R., et al., 2019, MNRAS, 490, 1
\bibitem[\protect\citeauthoryear{Bannister et al.}{2017}]{bannister17} 
Bannister, K. W., Shannon, R. M., et al., 2017, ApJ Letters, 841, L12
\bibitem[\protect\citeauthoryear{Bannister et al.}{2019}]{bannister19}
Bannister, K. W., Deller A. T., ey al., 2019, Science, 365, 565
\bibitem[\protect\citeauthoryear{Bera et al.}{2016}]{bera16} 
Bera A., Bhattacharyya S., et al., 2016, MNRAS, 457, 2530
\bibitem[\protect\citeauthoryear{Bhandari \& Keane}{2017}]{bhandari17} 
Bhandari, S., Keane, E. F., 2017, ArXiv e-prints: 1711.08110
\bibitem[\protect\citeauthoryear{Bhandari et al.}{2018}]{bhandari18} 
Bhandari, S., Caleb, M., et al., 2018, ATel: 12060
\bibitem[\protect\citeauthoryear{Bhandari et al.}{2019}]{bhandari19} 
Bhandari, S.; Bannister, K. W., et al., 2019, MNRAS, 486, 70
\bibitem[\protect\citeauthoryear{Bhat et al.}{2004}]{bhat04} 
Bhat N. D. R., Cordes J. M., et al., 2004, ApJ, 605, 759.
\bibitem[\protect\citeauthoryear{Boyle et al.}{2018}]{boyle18} 
Boyle P. J., et al., 2018, ATel: 11901
\bibitem[\protect\citeauthoryear{Caleb et al.}{2017}]{caleb17} 
Caleb, M., Flynn, C., et al., 2017, MNRAS, 468, 3746
\bibitem[\protect\citeauthoryear{Caleb et al.}{2018}]{caleb18}
Caleb, M., Spitler, L. G., et al. 2018, Nature Astronomy, 2, 839
\bibitem[\protect\citeauthoryear{Chatterjee et al.}{2017}]{chatterjee17}
Chatterjee, S., Law, C. J., et al., 2017, Nature, 541, 58
\bibitem[\protect\citeauthoryear{CHIME/FRB collaboration et al.}{2019a}]{chime19a} 
CHIME/FRB Collaboration et al., 2019a, Nature, 566, 230
\bibitem[\protect\citeauthoryear{CHIME/FRB collaboration et al.}{2019b}]{chime19b} 
CHIME/FRB Collaboration et al., 2019b, Nature, 566, 235
\bibitem[\protect\citeauthoryear{CHIME/FRB collaboration et al.}{2019c}]{chime19c} 
CHIME/FRB Collaboration et al., 2019c, ApJ Letters, 885, L24
\bibitem[\protect\citeauthoryear{CHIME/FRB collaboration et al.}{2020}]{chime20} 
CHIME/FRB Collaboration et al., 2020 ArXiv e-prints: 2005.10324
\bibitem[\protect\citeauthoryear{Conover}{1999}]{conover99} 
Conover, W. J., "Practical Nonparametric Statistics." John Wiley \& Sons, INC., 3rd Edition, 1999, ch. 6
\bibitem[\protect\citeauthoryear{Cordes \& Lazio}{2002}]{cordes02} 
Cordes J. M., Lazio T. J. W., 2002, ArXiv e-prints: astro-ph/0207156
\bibitem[\protect\citeauthoryear{Cordes et al.}{2016}]{cordes16} 
Cordes, J. M., Wharton, R. S., et al., 2016, ArXiv e-prints: 1605.05890
\bibitem[\protect\citeauthoryear{Farah et al.}{2018a}]{farah18a} 
Farah, W., Flynn, C., et al., 2018a, MNRAS, 478, 1209
\bibitem[\protect\citeauthoryear{Farah et al.}{2018b}]{farah18b} 
Farah, W., Bailes, M., et al., 2018b, ATel: 11675
\bibitem[\protect\citeauthoryear{Farah et al.}{2018c}]{farah18c} 
Farah, W., Bailes, M., et al., 2018c, ATel: 12335
\bibitem[\protect\citeauthoryear{Farah et al.}{2019}]{farah19} 
Farah, W., Flynn, C., et al., 2019, ArXiv e-prints: 1905.02293
\bibitem[\protect\citeauthoryear{Fedorova \& Rodin}{2019}]{fedorova19a} 
Fedorova, V. A., Rodin, A. E., 2019, Astronomy Reports, 63, 877
\bibitem[\protect\citeauthoryear{Hardy et al.}{2017}]{hardy17} 
Hardy, L. K., Dhillon, V. S., et al., 2017, MNRAS, 472, 2800 
\bibitem[\protect\citeauthoryear{Houben et al.}{2019}]{houben19} 
Houben, L. J. M., Spitler, L. G., et al., 2019, ArXiv e-prints: 1902.01779
\bibitem[\protect\citeauthoryear{Ioka}{2003}]{ioka03}
Ioka K., 2003, ApJ, 598, L79
\bibitem[\protect\citeauthoryear{James et al.}{2019}]{james19} 
James, C. W., Ekers, R. D., et al., 2018, MNRAS, 483, 1342
\bibitem[\protect\citeauthoryear{Keane et al.}{2016}]{keane16} 
Keane, E. F., Johnston, S., et al., 2016, Nature, 530, 453
\bibitem[\protect\citeauthoryear{Kendall \& Stuart}{1946}]{kendall46} 
Kendall, M. G., Stuart, A., 1946, The Advanced Theory of Statistics, Vol. 2, Griffin, London. 
\bibitem[\protect\citeauthoryear{Li et al.}{2020}]{li20} 
Li C.K., Lin L., et al., 2020, ArXiv e-prints: 2005.11071 
\bibitem[\protect\citeauthoryear{Lorimer et al.}{2007}]{lorimer07} 
Lorimer, D. R., Bailes, M., et al., 2007, Science, 318, 777
\bibitem[\protect\citeauthoryear{Lu \& Piro}{2019}]{lu19} 
Lu, W., Piro, A. L., 2019, ArXiv e-prints: 1903.00014.
\bibitem[\protect\citeauthoryear{Macquart \& Koay}{2013}]{macquart13} 
Macquart, J. P., Koay, J. Y., 2013, ApJ, 776, 125.
\bibitem[\protect\citeauthoryear{Macquart et al.}{2018}]{macquart18} 
Macquart, J. P., Shannon, R. M., et al., 2018 ArXiv e-prints: 1810.04353
\bibitem[\protect\citeauthoryear{Macquart et al.}{2019}]{macquart19} 
Macquart, J. P., Shannon, R. M., 2019, ApJ, 872, L19.
\bibitem[\protect\citeauthoryear{Macquart et al.}{2020}]{macquart20} 
Macquart, J. P., Prochaska J. X., et al., 2020, Nature, 581, 391
\bibitem[\protect\citeauthoryear{Madau \& Dickinson}{2014}]{madau14}
Madau P., Dickinson M., 2014, ARA\&A, 52, 415
\bibitem[\protect\citeauthoryear{Marcote et al.}{2020}]{marcote20} 
Marcote, B., Nimmo, K., et al., 2020, Nature, 577, 190
\bibitem[\protect\citeauthoryear{Masui et al.}{2015}]{masui15}
Masui, K., Hsiu-Hsien, L., et al., 2015, Nature, 528, 7583
\bibitem[\protect\citeauthoryear{Oslowski et al.}{2018a}]{osolowski18a} 
Oslowski, S., Shannon, R. M., et al., 2018a, ATel: 11385
\bibitem[\protect\citeauthoryear{Oslowski et al.}{2018b}]{osolowski18b} 
Oslowski, S., Shannon, R. M., et al., 2018b, ATel: 11396
\bibitem[\protect\citeauthoryear{Oslowski et al.}{2018c}]{osolowski18c} 
Oslowski, S., Shannon, R. M., et al., 2018c, ATel: 11851
\bibitem[\protect\citeauthoryear{Palaniswamy et al.}{2018}]{Palaniswamy18} 
Palaniswamy, D., Li, Y., et al., 2018, ApJ Letters, 854, L12
\bibitem[\protect\citeauthoryear{Patel et al.}{2018}]{patel18} 
Patel, C., Agarwal, D., et al., 2018, ArXiv e-prints: 1808.03710
\bibitem[\protect\citeauthoryear{Peacock}{1983}]{peacock83} 
Peacock, J. A., 1983, MNRAS, 202, 615 
\bibitem[\protect\citeauthoryear{Petroff et al.}{2016}]{petroff16} 
Petroff, E., Barr, E. D., et al. 2016, PASA, 33, 45 
\bibitem[\protect\citeauthoryear{Petroff et al.}{2017}]{petroff17} 
Petroff, E., Burke-Spolaor, S., et al., 2017, MNRAS, 469, 4465
\bibitem[\protect\citeauthoryear{Planck collaboration et al.}{2013}]{planck13}
Planck Collaboration et al., 2013, A\&A, 571, 16
\bibitem[\protect\citeauthoryear{Platts et al.}{2018}]{platts18} 
Platts, E., Weltman, A., et al., 2018, ArXiv e-prints: 1810.05836
\bibitem[\protect\citeauthoryear{Price et al.}{2018}]{price18} 
Price, D. C., Gajjar, V., et al., 2018, 2018, ATel: 11376
\bibitem[\protect\citeauthoryear{Prochaska et al.}{2019}]{prochaska19} 
Prochaska, J. X., Macquart, J. P., et al. 2019, Science, 366, 231
\bibitem[\protect\citeauthoryear{Ravi et al.}{2016}]{ravi16} 
Ravi, V., Shannon, R. M., et al., 2016, Science, 354, 6317
\bibitem[\protect\citeauthoryear{Ravi \& Loeb}{2018}]{ravi18} 
Ravi, V., Loeb, A., 2018, ArXiv e-prints: 1811.00109
\bibitem[\protect\citeauthoryear{Ravi et al.}{2019}]{ravi19}
Ravi, V., Catha, M., et al. 2019, Nature, 572, 352
\bibitem[\protect\citeauthoryear{Ridnaia et al.}{2020}]{ridnaia20} 
Ridnaia A., Svinkin D., et al., 2020, ArXiv e-prints: 2005.11178
\bibitem[\protect\citeauthoryear{Rowlinson et al.}{2016}]{rowlinson16} 
Rowlinson, A., Bell, M. E., et al., 2016, MNRAS, 458, 3506.
\bibitem[\protect\citeauthoryear{Schechter}{1976}]{schechter76}
Schechter, P., 1976, ApJ, 203, 297. 
\bibitem[\protect\citeauthoryear{Shannon et al.}{2017}]{shannon17} 
Shannon, R. M., Oslowski, S., et al., 2017, ATel: 11046
\bibitem[\protect\citeauthoryear{Shannon et al.}{2018}]{shannon18} 
Shannon, R. M., Macquart, J. P., et al., 2018, Nature, 562, 386
\bibitem[\protect\citeauthoryear{Sokolowski et al.}{2018}]{sokolowski18}
Sokolowski M., Bhat, N. D. R., et al., 2018, ApJ, 867, L12. 
\bibitem[\protect\citeauthoryear{Spitler et al.}{2014}]{spitler14} 
Spitler L. G., Cordes, J. M., et al., 2014, ApJ, 790, 101
\bibitem[\protect\citeauthoryear{Tavani et al.}{2020}]{tavani20} 
Tavani M., Casentini C., et al., 2020, ArXiv e-prints: 2005.12164
\bibitem[\protect\citeauthoryear{Thornton et al.}{2013}]{thornton13} 
Thornton, D., Stappers, B., et al., 2013, Science, 341, 53
\bibitem[\protect\citeauthoryear{Qiu et al.}{2019}]{qiu19} 
Qiu, Hao, Bannister, K. W., et al., 2019, MNRAS, 486, 166

\end{thebibliography}
\end{document}